\documentclass[showpacs,prd,superscriptaddress,nofootinbib,twocolumn]{revtex4}
\usepackage{amsmath,amssymb,epsfig,natbib}
\def\bea{\begin{eqnarray}}
\def\eea{\end{eqnarray}}
\def\S{{\cal{S}}}
\newcommand\fig[1]{Fig.~\ref{#1}}
\renewcommand\sec[1]{Sec.~\ref{#1}}

\newcommand{\fnl}{f_{\rm nl}}

\newcommand{\zetarad}{\zeta^{\text{rad}}}

\newcommand\ba{\begin{eqnarray}}
\newcommand\ea{\end{eqnarray}}
\newcommand\be{\begin{equation}}
\newcommand\ee{\end{equation}}

\newcommand\del{\nabla}

\newcommand{\cdm}{{\rm cdm}}
\newcommand{\B}{\mathrm B}
\renewcommand{\L}{\mathrm L}

\renewcommand{\S}{\mbox{$\cal S$}}

\newcommand\eq[1]{Eq.~(\ref{#1})}
\newcommand\eqs[2]{Eqs.~(\ref{#1}) and (\ref{#2})}
\newcommand\eqss[3]{Eqs.~(\ref{#1}), (\ref{#2}) and (\ref{#3})}

\newcommand\slabel[1]{\label{#1}}

\begin{document}
%
\title{WMAP, neutrino degeneracy and non-Gaussianity constraints 
on isocurvature perturbations in the curvaton model 
of inflation}

\author{Christopher Gordon}
\email{cgordon@cfcp.uchicago.edu}
\affiliation{Enrico Fermi Institute and Center for Cosmological Physics,
  University of Chicago, USA \\
  Department of Applied Mathematics and Theoretical Physics, University
 of Cambridge, UK.}
\author{Karim A.~Malik}
\email{k.malik@lancaster.ac.uk}
\affiliation{
Physics Department, Lancaster University, Lancaster LA1 4YB, UK
\\
GRECO, Institut d'Astrophysique de Paris, C.N.R.S.,
75014 Paris, France.
}
%
\begin{abstract}
 In the curvaton model of inflation, where a second scalar field, the
`curvaton', is responsible for the observed inhomogeneity, a non-zero
neutrino degeneracy may lead to a characteristic pattern of
isocurvature perturbations in the neutrino, cold dark matter and
baryon components. We find the current data can only place upper
limits on the level of isocurvature perturbations. These can be
translated into upper limits on the neutrino degeneracy parameter. In
the case that lepton number is created before curvaton decay, we find
that the limit on the neutrino degeneracy parameter is comparable with that
obtained from Big-bang nucleosynthesis. For the case that lepton
number is created by curvaton decay we find that the absolute value
of the non-Gaussianity parameter,$|\fnl|$, must be less than 10 (95\%
confidence interval).
\end{abstract}

\pacs{98.80.Cq}
\maketitle

\section{Introduction}

In single field models  of 
inflation,
the inflaton
gives rise to an era of accelerated expansion in the
early universe during which quantum fluctuations in the inflaton field
are expanded beyond the horizon size \cite{LLBook}. At the end of
inflation, the inflaton decays into the standard matter fields.  The
primordial fluctuations, imprinted onto the gravitational potential in
form of small perturbations, become the ``seeds'' that start the
formation of large scale structure. As there is only a single degree
of freedom, the inflaton's value, the resulting perturbations are
always {\em adiabatic,} i.e.\ the ratio of the number densities of
different particle species is spatially homogeneous, there is no
isocurvature mode.

Recently a different model has been proposed, the curvaton scenario 
\cite{curvaton, LUW}, which has already received considerable 
attention \cite{allcurvaton,MWU,DHLDW,gordonlewis}. 
In this model the accelerated expansion is still provided by the 
inflaton, but the primordial perturbations are generated
by another scalar field, the {\em curvaton}.
The curvaton energy density is small compared to the inflaton 
energy density and being
a light field, it acquires an almost scale-invariant spectrum 
of perturbations during inflation. After
the decay of the inflaton, the isocurvature perturbation in the
curvaton fluid is transformed into an adiabatic perturbation
\cite{Mollerach,LM,MWU} and the curvaton decays into radiation and other
particles. A similar model to the curvaton scenario, in which
a second scalar field determines the
inflaton decay rate, has also been  proposed recently \cite{vardecay}.

The curvaton scenario requires different constraints on the
inflationary potential to standard inflation and might therefore be
better suited for inflation at a lower energy scale
\cite{lythdimopolos,lythsmallscale}. It also allows for the
possibility of an isocurvature mode in the standard matter 
fields
\cite{curvaton,LUW}, i.e.\ the ratio of particle number densities of
different particle species is not necessarily spatially
homogeneous. This arises since the final constituents of the
standard big bang model (photons, neutrinos, baryons and cold dark
matter (CDM)) do not all necessarily originate entirely from the
spatially inhomogeneous curvaton, but may originate instead from the
spatially homogeneous inflaton or some other source.

In Ref.~\cite{gordonlewis} the
isocurvature components in the CDM and baryons that could arise in the
curvaton scenario were constrained using WMAP and other observational
data. It was found that the current data rule out either CDM or
baryons being created before curvaton decay. In this article we
examine what the current observational constraints are on neutrino
isocurvature perturbations created by curvaton decay. These can only
arise if there is a non-zero neutrino degeneracy parameter $\xi =
\mu_\nu / T_\nu$, where $\mu_\nu$ is the neutrino chemical potential
and $T_\nu$ is the neutrino temperature \cite{LUW}.
There may be a separate degeneracy parameter for each of
the three species of neutrinos, but the recent atmospheric and solar
neutrino data indicate that approximate flavor equilibrium between all active
neutrino species is established well before the BBN epoch \cite{solar}
and hence we take the degeneracy parameters to be equal.

Various studies have constrained the value of $\xi$ using
CMB and other large scale structure data in the adiabatic case
\cite{xibg}. However, the tightest constraints still come from 
Big-bang nucleosynthesis (BBN) \cite{bbnxi}. In this article we examine
how the constraints on $\xi$ change in the curvaton scenario where
neutrino isocurvature perturbations can be induced by a non-zero
$\xi$. It turns out that the neutrino isocurvature perturbations then
also induce baryon and CDM isocurvature perturbations \cite{LUW}. 
It is therefore not possible to use constraints that have simply considered the
addition of a neutrino isocurvature mode such as
Ref.~\cite{garcia-bellido} or those that have considered arbitrary
combinations of different isocurvature perturbations \cite{multiiso}.

In the next section we evaluate the possible magnitudes of the
neutrino isocurvature perturbations. Then in \sec{section3} we
estimate analytically the effects of the isocurvature perturbations on
the temperature fluctuations of the CMB and their dependence on
$\xi$. A full likelihood analysis of the allowed ranges of $\xi$ is
given in \sec{data}. Finally, the conclusions and the discussion are given in
\sec{conclusions}.

\section{Curvaton generated neutrino isocurvature perturbations}
\label{section2}

A useful gauge-invariant variable for characterizing inhomogeneity on
large scales is \cite{BST,Bardeen88,WMLL} 
\be \zeta = -\psi - H
\frac{\delta\rho}{\dot\rho}\,, 
\slabel{zetadef} 
\ee
where $\rho$ and $H$ are respectively the total energy density and
the Hubble parameter, the dot denotes differentiation with respect
to coordinate time, and the
perturbation in a background quantity $x$ is denoted by $\delta x$.
For a critical density Universe, $\psi$ is related to the intrinsic 
curvature of a spatial
hypersurface by \cite{WMLL}
\be
^{(3)}R = \frac{4}{a^2} \del^2 \psi \,,
\ee
where $a$ is the scale factor and $\del_i$ is the covariant
derivative. 
It is possible to choose the coordinate
system such that certain combinations of perturbation variables are
set to zero. This is known as choosing the gauge or frame
\cite{Bardeen,KS,WMLL}. In this article we will work in the {\em flat\/} gauge
\cite{Bardeen,KS,WMLL} in which $\psi=0$, and \eq{zetadef} becomes
\be 
\zeta = -H\frac{\delta\rho}{\dot\rho}\,,
\slabel{zeta}
\ee
where from now on $\delta x$ refers to the perturbation in $x$ when the flat
gauge is chosen. The energy conservation equation is given by
\be
\dot\rho+3H(\rho+P)=0\,,
\slabel{dotrho}
\ee
where $P$ is the total pressure.
After curvaton ($\sigma$) decay, the component fluids are
photons ($\gamma$), neutrinos $(\nu)$, cold dark matter ($\cdm$) and
baryons ($\B$). The total energy density and pressure are 
related to density and pressure  of a component 
$i$ by $\rho=\sum_i \rho_i$ and $P=\sum_i P_i$. 
 The component pressures are given in terms of the corresponding
 component densities:
$P_i=\rho_i/3$ for photons and neutrinos and $P_i=0$ for the curvaton
(after the end of inflation), CDM and baryons.

We assume a critical density Universe and that the neutrinos are 
effectively massless.

The isocurvature perturbations are given by \cite{MWU}
\be \S_i = 3(\zeta_i - \zeta_\gamma) \,,
\slabel{Si} 
\ee
where $i$ is one of $\sigma$, $\nu$, $\cdm$ or $\B$ and 
\be 
\zeta_i
= -H\frac{\delta\rho_i}{\dot\rho_i} \,.
\slabel{zetai} 
\ee
If there is no energy transfer to and from a fluid so that it satisfies the energy
conservation equation:
\be
\dot{\rho_i} + 3H(\rho_i + P_i) = 0\,,
\ee
then
 its $\zeta_i$ is conserved on large scales,
$\dot{\zeta_i}=0$ \cite{WMLL}.

The occupation number for a neutrino species
($i=e$, $\mu$ or $\tau$) with energy $E$ is given by 
\be f_i(E) =
[\exp(E/T_\nu \mp \xi_i) + 1]^{-1} \,,
\ee 
where $T_\nu$ is the neutrino temperature and the minus is for
neutrinos and the plus for anti-neutrinos. The degeneracy parameter is
$\xi_i = \mu_i /T_\nu$ where $\mu_i$ is the chemical potential of
species $i$. However, using the Large Mixing Angle (LMA) 
solution leads to the degeneracy parameter
being approximately the same for all three species \cite{solar}, i.e.\
$\xi_i=\xi$. After positron annihilation, the asymmetry parameter,
$\xi$, is constant in time \cite{LUW}. When $\xi$ is non-zero, the
difference between the number densities of neutrinos and anti-neutrinos
is non-zero. We denote this quantity $n_L$ as this difference is equal
to the lepton number. Given the
constraints on charge asymmetry,
any non-negligible
lepton number must be due to the neutrinos.
There can be no neutrino isocurvature
perturbation when $n_L=0$ as then the neutrino energy density is
determined solely by the photon energy density before neutrino
decoupling \cite{LUW}.

Analogously to the case of energy density, \eq{zetai}, the
inhomogeneity of lepton number can be characterized by 
\be
{\zeta}_L = -H\frac{\delta n_L}{\dot n_L} \,,
\label{defdeltanumber}
\ee
where as specified before we are using $\delta n_L$ in the flat gauge.
If the lepton number density is conserved
\be
\dot{n}_L + 3H n_L = 0 \,,
\ee
then $\dot{\zeta}_L = 0$ on large scales \cite{LUW,DHLDW}.

The isocurvature perturbation of the lepton
number perturbation is then given by 
\be
{\S}_\L = 3({\zeta}_\L - \zeta_\gamma)\,,
\slabel{tildeS0}
\ee
which is related to the neutrino isocurvature perturbation by \cite{LUW}
\be
\S_\nu = \frac{45}{7}\frac{B^2}{B'A}\S_\L\,,
\slabel{Snu0}
\ee
where 
\begin{eqnarray}
A &=& \left[3.04/3 + \frac{30}{7} \left(\frac{\xi}{\pi}\right)^2
 + \frac{15}{7} \left(\frac{\xi}{\pi} \right)^4 \right]\,,\\
B &=& \left[ \frac{\xi}{\pi} + \left(\frac{\xi}{\pi} \right)^3 \right]\,, 
 \slabel{A}
\end{eqnarray}
and $B' = 1 + 3(\xi/\pi)^2$.  
The $3.04/3$ term in the definition of $A$ takes into account the
non-equilibrium heating of the neutrino fluid and finite temperature
QED corrections \cite{Dolgov,bbnxi,QED}.

If the lepton number is created well before the curvaton decay then
it will have negligible perturbations 
(${\zeta_\L} =0$) and so from \eq{tildeS0}
\be
{\S}_\L = -3\zeta_\gamma \,.
\slabel{tildeS}
\ee
We will not make the assumption $|\xi| \ll 1$, which was used in 
Ref.~\cite{curvaton} as we will be also considering the case where $\xi$
may be large. To evaluate the effect of $\S_\nu$ on
observations we need to express it in terms of the adiabatic
perturbation $\zeta$. To do this we can use \eqss{zeta}{dotrho}{zetai} and that
only the
photons and neutrinos contribute significantly to the density in the
primordial era to get 
\be
\zeta = (1-R_\nu) \zeta_\gamma + R_\nu \zeta_\nu\,,
\slabel{zetaRnu}
\ee
where 
\be
R_\nu = \frac{\rho_\nu}{\rho_\nu + \rho_\gamma}\,,
\slabel{Rnu}
\ee
which can be evaluated using \cite{Dolgov,bbnxi}
\be
\rho_\nu =  \frac{7}{8} \left( \frac{4}{11} \right)^{4/3} 3A
\rho_\gamma \,.
\slabel{rhonu}
\ee
Combining \eq{zetaRnu} with \eq{Si} for $i=\nu$ we get
\be
\zeta = \zeta_\gamma + \frac{1}{3} R_\nu \S_\nu\,,
\slabel{zeta1}
\ee
then solving \eq{zeta1} for $\zeta_\gamma$ and substituting 
into \eq{tildeS} 
gives 
\be
\S_L = -3\zeta + R_\nu \S_\nu.
\slabel{tildeS2}
\ee
Substituting \eq{tildeS2} into \eq{Snu0} and solving for the
 neutrino isocurvature perturbation gives
\be
\S_\nu = \frac{135 B^2}{45 B^2 R_\nu-7B'A}\zeta\,.
\slabel{Snubefore}
\ee
If the lepton number is created by curvaton decay then \cite{LUW}
\be
{\zeta}_L = \zeta_\sigma =  \frac{\zeta}{r} \,,
\slabel{zetaL}
\ee
where  using the sudden decay approximation, 
$r \approx \rho_\sigma / (\rho_\sigma + \rho_\gamma)$ evaluated at the
time of curvaton decay. The 
approximation is accurate up to about 10
\% error even when the decay is not sudden \cite{MWU}. 
Using \eq{Si} (with $i=\nu$), \eqss{tildeS0}{zetaRnu}{zetaL} gives
\be
\S_L = 3\frac{1-r}{r} \zeta+R_\nu\S_\nu \,.
\slabel{tildeS1}
\ee
Substituting \eq{tildeS1} into \eq{Snu0} and solving for $\S_\nu$ we get
\be
\S_\nu=\frac{r-1}{r}\frac{135 B^2}{45B^2 R_\nu-7 B'A}\zeta\,,
\slabel{Snu1}
\ee
which is valid even when $\xi^2$ is large. As can be seen, the
neutrino isocurvature perturbation resulting from lepton number
created by curvaton decay, \eq{Snu1}, differs from the neutrino
isocurvature perturbation result when lepton number is created before
curvaton decay, \eq{Snubefore}, by a factor which is only a function
of $r$.

The final possibility is for the lepton number to be created after
curvaton decay which corresponds to $r=1$ in \eq{Snu1} and so in this case
there is no neutrino isocurvature perturbation.

If there is a neutrino isocurvature perturbation it induces a
CDM and baryon isocurvature perturbation \cite{LUW}. In the case where
CDM or baryons are created after curvaton decay \cite{LUW}:
\be
\S_{\cdm/\B} = R_\nu \S_\nu\,,
\slabel{induced}
\ee
where $\S_{\cdm/\B}$ stands for $\S_{\cdm} $ or $\S_\B$ as the effect is
the same on either one. When CDM/B is created before curvaton decay
there is an additional residual contribution \cite{LUW}:
\be
\S_{\cdm/\B} = R_\nu \S_\nu - 3 \zeta\,.
\ee
When CDM/B is created by curvaton decay
\be
\S_{\cdm/\B} = R_\nu \S_\nu + 3\frac{1-r}{r} \zeta\,.
\ee
However,  to concentrate on the role of neutrino isocurvature
perturbations we will only consider the case where baryon number and CDM are
created after curvaton decay, \eq{induced}.

\section{Neutrino Degeneracy Modifications of the Sachs-Wolfe effect}
\label{section3}

In this section we examine analytically the effect of a neutrino
degeneracy in the curvaton scenario.

In the presence of isocurvature perturbations the temperature
fluctuations due to the Sachs-Wolfe effect \cite{SW,KSrad1} on large
scales are given
by \cite{gordonlewis}
\be 
\frac{\Delta T}{T} = -\frac{1}{5}\zetarad - \frac{2}{5} \left( R_\B \S_\B +
R_\cdm \S_{\cdm} \right) + \frac{1}{15} R_\nu \S_\nu \,,
\slabel{SW}
\ee
where $\zetarad$ is the value of $\zeta$ in the radiation
era and
\begin{eqnarray}
R_\B &=& \frac{\rho_\B}{\rho_\B + \rho_\cdm}, \slabel{RB}\\
R_\cdm &=& \frac{\rho_\cdm}{\rho_\B + \rho_\cdm}. \slabel{Rcdm}
\end{eqnarray}
Substituting the induced baryon isocurvature perturbations
\eq{induced} into \eq{SW} and making use of \eqs{RB}{Rcdm} gives
\be
\frac{\Delta T}{T}=-\frac{1}{5}\zetarad -\frac{1}{3}R_\nu \S_\nu.
\slabel{SW1}
\ee
The current 95\% confidence interval (CI) limits on $\xi$ from 
BBN are  \cite{bbnxi}
\be
-0.03 \le \xi \le 0.11.
\slabel{bbnxi}
\ee
 So in order to determine
whether within these limits the isocurvature perturbations can still
play a role we can use $\xi\ll 1$ which gives from
\eqss{A}{Rnu}{rhonu} gives
\be
R_\nu \approx 0.41.
\slabel{Rnu2}
\ee
Substituting \eq{Rnu2} into \eq{SW1} gives
\be
\frac{\Delta T}{T}\approx-\frac{1}{5} \left(\zetarad +\frac{1}{2}
\S_\nu \right) \,,
\slabel{SW2}
\ee
which shows that the neutrino isocurvature perturbation has a similar
sized effect as the adiabatic perturbation on large scales.

For the lepton number created before curvaton decay with $\xi \ll 1$,
\eq{Snubefore} gives
\be
\S_\nu \approx -\frac{135}{7} \left(\frac{\xi}{\pi} \right)^2 \zeta
\slabel{Snubefore1}
\ee
which when substituted into \eq{SW2} results in
\be
\frac{\Delta T}{T} \approx \frac{1}{5} \left( -1  +
\xi^2 \right) \zetarad.
\slabel{SWbefore}
\ee
Hence in this case the neutrino degeneracy decreases the temperature
fluctuations on large scales. However, for the BBN limits of $|\xi| \le
0.11$ it follows from \eq{SWbefore} that the isocurvature
perturbation has a negligible effect on the temperature fluctuations
when the lepton number is created before curvaton decay.

For lepton number created by curvaton decay with $\xi \ll 1$,
\eq{Snu1} gives
\be
\S_\nu = \frac{1-r}{r} \frac{135}{7} \left(\frac{\xi}{\pi} \right)^2 \zeta.
\slabel{Snuby}
\ee
>From \eq{Snuby} and \eq{SW2} we then find
\be
\frac{\Delta T}{T} \approx \frac{1}{5}  \left( -1 +  \frac{r-1}{r}
\xi^2 \right) \zetarad\,.
\slabel{SWby}
\ee
As can be seen from \eq{SWby}, in the case where the lepton number is
created by curvaton decay, a non-zero $\xi$ adds to the magnitude of
the fluctuations on large scales as $0 \le r \le 1$. Limits on $r$ can be
placed using the limits on the non-Gaussianity parameter $\fnl$
\cite{LUW,curvatonfnl}
\be
\fnl = \frac{1}{3} + \frac{5}{6} r - \frac{5}{4 r}\,.
\slabel{fnl}
\ee
The current WMAP limits two sigma confidence interval is \cite{Komatsu}
\be
-58 \le \fnl \le 134\,.
\ee

Inverting \eq{fnl} this leads to $r \ge 0.02$ at two sigma confidence
level. So from \eq{SWby}, it appears that there may be scope for a
large contribution by the isocurvature term. We will test this with a
full likelihood approach in the next Section.

\section{Data Analysis}
\slabel{data}
The effect of a non-zero $\xi$ on all cosmological scales
can be obtained
by
inputting the isocurvature perturbations derived in
\eqs{Snubefore}{Snu1} and the induced baryon and CDM isocurvature
perturbations, \eq{induced}, into CAMB\footnote{http://camb.info/}. To account for the
effect on the background dynamics the number of massless neutrinos
needs to be 
set to 
\be
N_\nu = 3A 
\slabel{Nnu}
\ee
where $A$ is defined in \eq{A}.
Some example cases are
plotted in \fig{cls}.
\begin{figure}
\begin{center}
\epsfig{figure=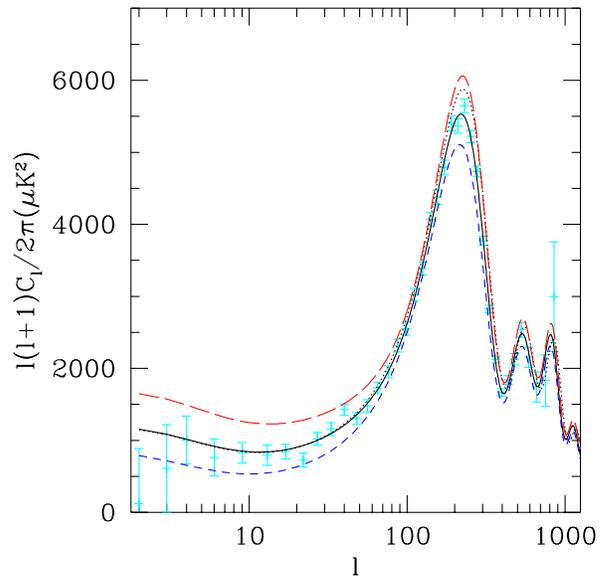,width=8cm}
\end{center}
\caption{\slabel{cls} 
An illustration of the effect of the lepton degeneracy ($\xi$) on the
spectrum of fluctuations. The cosmological parameters (including the
amplitude) for all the spectra are taken from the maximum likelihood
when $\xi=0$. The solid line is a plot of the pure adiabatic
spectrum. The short dashed line is for the case when the lepton number
is created before curvaton decay and $\xi=0.3$. The long dashed line is
for the case when the lepton number is created by curvaton decay with
$\xi=0.3$ and the fraction of the density made up by the curvaton at
curvaton decay ($r$) is set to 0.5. 
The dotted line is the spectrum for when the lepton number is created
after curvaton decay (so that there are no isocurvature modes) and $\xi=1$.
The binned WMAP TT data is also plotted.}
\end{figure}
This figure confirms the direction of the effect evaluated
analytically on large scales in \eqs{SWbefore}{SWby}, i.e.\ a non-zero
$\xi$ decreases the power on large scales when the lepton number is
created before curvaton decay and increases it when the lepton number
is created by curvaton decay. As can be see from \fig{cls}, there is a
similar effect on the extrema of the acoustic peaks but {\em not\/}
on the rise and fall of the acoustic peaks. There is also a slight
shift in the phase of the oscillations. As can also be seen in
\fig{cls}, a similar effect of a non-zero
$\xi$ occurs 
when there are no isocurvature modes \cite{xibg,bashinskyseljak}
but then there is no effect on
the Sachs-Wolfe plateau.
 Note that the decrease or increase is fairly constant
over large scales and so is not suitable for resolving the quadrupole
problem \cite{quad}.

The current BBN constraints on $\xi$ are given in \eq{bbnxi}. They rely
on using the large angle solution in which all the $\xi$ for the
three neutrino species become equal \cite{solar}. The actual constraint is on
$\xi_e$, the electron neutrino degeneracy, due to its effect on the neutron to
proton ratio prior to BBN. There have been attempts to check
the BBN constraint by independently using CMB and other large scale
structure data \cite{xibg}. This has been done in the context of
adiabatic perturbations and so the effect is due to the increase of
density caused by an greater effective number of neutrinos,
\eq{Nnu}. 

In this Section we test what the constraints are when lepton number is
created before and by curvaton decay. The adiabatic case is equivalent
to when lepton number is created after curvaton decay as it
corresponds to setting $r=1$ in \eq{Snu1}.  A modified version of
COSMOMC\footnote{http://cosmologist.info/cosmomc} \cite{cosmomc} was
used for the likelihood analysis which included WMAP temperature and
temperature-polarization cross-correlation anisotropy \cite{WMAP},
ACBAR \cite{ACBAR}, CBI \cite{CBI}, 2dF galaxy redshift survey
\cite{2dF} and Hubble Space Telescope (HST) Key Project \cite{HST}
data.  The Universe was taken as flat and the neutrinos as
massless. Reionization was parameterized using optical depth
\cite{Kosowsky}.  Each simulation used five separate Markov chains and
the convergence criteria was taken to be the variance of the chain mean divided
by the mean of the chain variance to be less than 0.2 for each
parameter.

As COSMOMC produces Monte Carlo samples of the output
variables, marginalization simply entails extracting the samples of
the variables of interest from the multi-dimensional samples. These
samples can then be used to evaluate confidence intervals or to give
estimates of the probability distribution by using
histograms. Functions of the variables can also be analyzed just by
taking functions of the corresponding samples. 

The different scenarios that were checked were: lepton number created
before curvaton decay, \eq{Snubefore}, lepton number created by
curvaton decay, \eq{Snu1}, 
 and
lepton number created after curvaton decay. COSMOMC was modified to
take extra variables $\xi$ and $r$ and their effect on the
isocurvature modes, \eqss{Snubefore}{Snu1}{induced}, 
and effective number of neutrinos, \eq{Nnu}, were also
added. Modifications were also made to allow the inclusion of a BBN
prior on $\xi$ which is a Gaussian distribution with mean 0.04 and
standard deviation $0.035$ \cite{bbnxi}. The prior for $\fnl$ was a
Gaussian distribution with a mean of 38 and standard deviation of 48
\cite{Komatsu}. This then imposed a prior on $r$ through \eq{fnl}.

In the cases where the BBN constraint was not used,
only positive values of $\xi$ were sampled as then the results were
even in $\xi$ as \eqss{Snubefore}{Snu1}{Nnu} are even in $\xi$. 

The constraints on $\xi$ for the various possibilities considered
are given in Table~\ref{results}. The marginalized distributions for
$\xi$ are also plotted in \fig{withandwithout}.

The BBN constraint is still tighter
than that obtained by using the CMB, 2dF and HST data on any of the
curvaton scenarios.

The case where lepton number is created before
curvaton decay is only twice as broad without the BBN
constraint.  Also, in this case,
the isocurvature contribution is almost 10\% as large as the adiabatic
contribution but this gets dramatically reduced when the BBN constraint
is added. 

The case when the lepton number is created by curvaton decay is less
constraining on $\xi$ even though 
the data do allow about as much positive isocurvature perturbations as
negative. This is because there is an additional degeneracy with $r$
as seen in \eq{Snu1} and so  also in the two dimensional marginalized
distribution of $\xi$ and $r$ as seen in \fig{xivsr}. Adding the BBN
constraint also dramatically decreases the amount of isocurvature
perturbations in this case as the smaller values of $r$ needed to
compensate for the lower $\xi$ is disfavored by the WMAP
non-Gaussianity constraints on $\fnl$. However, smaller values of
$r$ are more likely with the BBN constraint and these lead to values of
$\fnl$ large enough in magnitude to possibly be detectable by 
the Planck satellite which may be sensitive to $|\fnl|>5$ \cite{planck}.

In the case where lepton number is created after curvaton decay, the
constraints on $\xi$ are the broadest. In this case there are no
isocurvature perturbations and a non-zero $\xi$ only effects the
background dynamics through modifying the effective neutrino number,
\eq{Nnu}. This case has been studied before 
  with similar results \cite{xibg}.

\begin{table}
\begin{tabular}{|l|r|r|r|}
\hline
Constraint & $\xi$ & $\S_\nu/\zetarad$ & $\fnl$ \\ \hline
BBN only & 0.1 & & \\
before curvaton decay   & 0.2  & -0.08 & \\
BBN + before curvaton decay & 0.1  & -0.02 &  \\
by curvaton decay &0.8 &  0.07  & -4  \\ 
BBN + by curvaton decay &  0.1 & 0.04 & -10   \\ 
after curvaton decay &  1.6 &  & \\ \hline
\end{tabular}
\caption{\slabel{results} 
Upper bound (95\% CI) constraints on the magnitudes of 
 the lepton degeneracy ($\xi$), the neutrino
isocurvature perturbation ($\S_\nu$) and the 
non-Gaussianity parameter $\fnl$.
The `BBN' label means  the constraint
from BBN using the large mixing angle solution \cite{bbnxi} is
included. The constraints from WMAP, ACBAR, CBI, 2dF and HST on the various
curvaton mechanisms of generating the lepton number are given.
}
\end{table}

\begin{figure}
\begin{center}
\epsfig{figure=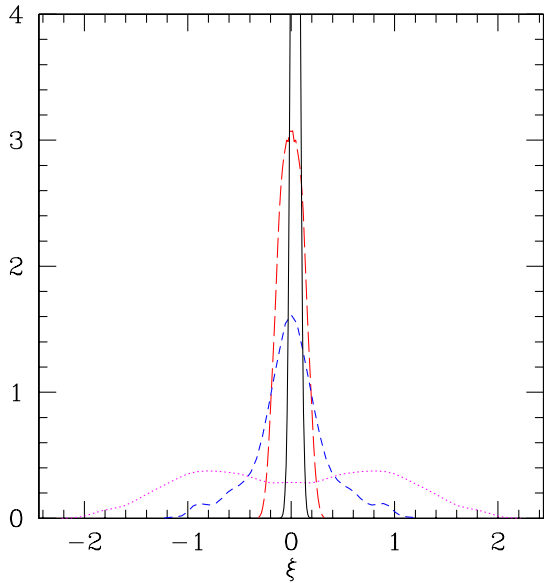,width=8cm}
\end{center}
\caption{\slabel{withandwithout} 
Comparison of marginalized probability distributions, 
using CMB, 2dF and HST data,
for 
 lepton degeneracy ($\xi$)  when lepton number is created by curvaton
decay (short-dashes), for lepton number created after curvaton decay
(dotted)
and for lepton number created before curvaton decay (long dashes).
The BBN constraint on $\xi$ is also plotted (solid line).
}
\end{figure}

\begin{figure}
\begin{center}
\epsfig{figure=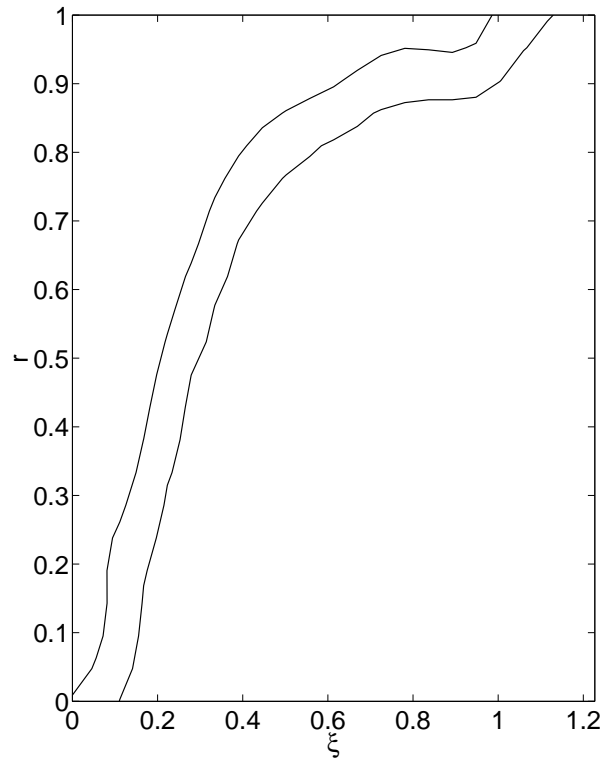,width=8cm}
\end{center}
\caption{ \slabel{xivsr} One and two sigma contours for the
2D marginalized probability distribution, using CMB, 2dF and HST data,
 of the lepton degeneracy ($\xi$) and the
curvaton density proportion at time of curvaton decay ($r$) for the
case when lepton number is created by curvaton decay.
}
\end{figure}

\section{Conclusions}
\slabel{conclusions}

We have examined the current cosmological constraints on the neutrino
degeneracy 
$\xi$, in the curvaton scenario, using WMAP, ACBAR, CBI, 2dF,
HST and BBN data. The expressions for the isocurvature perturbations were
evaluated to all orders in $\xi$ and the effects of the induced
baryon and CDM isocurvature perturbations were accounted
for. Analytical formulas were given for the modification of the
Sachs-Wolfe effect on large scales.

Comparisons were made with the BBN constraint which results from the
effect of the neutrino degeneracy on the pre-BBN proton to neutron
ratio and using the large mixing angle solution in which the neutrino
degeneracies of the different species take on the same value.  It was
found that the constraint on $\xi$ was comparable to that from BBN
when the lepton number was created before curvaton decay. 
As seen in \fig{cls}, a non-zero $\xi$ affects the
acoustic peaks whose data will dramatically improve in the future. So
the CMB constraints on $\xi$ may also improve dramatically.
 Without the BBN constraint, the neutrino isocurvature
perturbation was found to be smaller than 10\% the size of the
adiabatic mode (95\% CI), while including the BBN constraint reduced
this by a factor of a few. In the case where the lepton number is
created by curvaton decay, there is a degeneracy between the value of
the curvaton energy density at the time of curvaton decay and $\xi$.
However, this does not stop the BBN data strengthening the
constraint on the isocurvature perturbation as the value of the
curvaton energy density at the time of curvaton decay is constrained
by the limits on the non-Gaussianity parameter from WMAP. So in this
case, the neutrino isocurvature mode is less than about 5\% (95\% CI)
of the adiabatic mode.
Unless the BBN constraints on $\xi$ become tighter around zero,
 the case of the lepton number being created by curvaton decay
would be unlikely if future observations found $\fnl < -10$, as this
would imply a larger isocurvature perturbation than is observed.

\acknowledgments

The authors are grateful to David Lyth and David Wands for useful
comments. CG thanks Antony Lewis for help with COSMOMC.  This work was
performed on COSMOS, the Origin3800 and Altix3700 owned by the UK
Computational Cosmology Consortium, supported by Silicon Graphics/Cray
Research, HEFCE and PPARC. CG was supported at the University of 
Cambridge by PPARC and
at the University of Chicago and at the Center for Cosmological
Physics by grant NSF PHY-0114422.
KM was supported by a Marie Curie
Fellowship under the contract number \emph{HPMF-CT-2000-00981} at IAP 
and by PPARC at the University of Lancaster.

{}

\end{document}